\documentstyle[11pt,newpasp,twoside,epsf]{article}
\def\edcomment#1{\iffalse\marginpar{\raggedright\sl#1\/}\else\relax\fi}
\reversemarginpar

\begin{document}
\title{Revised analysis of the supersoft X-ray phase, helium enrichment,
and turn-off time in the 2000 outburst of the recurrent nova CI Aquilae
} 

\author{Izumi Hachisu} 

\affil{College of Arts and Sciences, University of Tokyo, 
Meguro-ku, Tokyo 153-8902, Japan}

\author{Mariko Kato} 

\affil{Keio University, Kouhoku-ku, Yokohama 223-8521, Japan}

\begin{abstract}
     The recurrent nova CI Aquilae entered the final decline phase
a bit before May of 2001, showing the slowest evolution
among the recurrent novae.  Based on the optically thick wind mass-loss
theory of the thermonuclear runaway model, we have estimated
the turn-off time of the CI Aql 2000 outburst in March of 2001,
after a supersoft X-ray source (SSS) phase lasts 150 days from
December of 2000 until May of 2001.  Fitting our theoretical 
light curves with both the 1917 and 2000 outbursts,
we also obtained the WD mass to be $M_{\rm WD}= 1.2 \pm 0.05 ~M_\odot$,
helium enrichment of ejecta, He/H$\sim 0.5$ by number,
the mass of the hydrogen-rich envelope on the WD of
$\Delta M \sim 6 \times 10^{-6} M_\odot$ at the optical maximum, 
which is indicating an average mass accretion rate of
$\dot M_{\rm acc} \sim 0.8 \times 10^{-7} M_\odot$ yr$^{-1}$ 
during the quiescent phase between the 1917 and 2000 outbursts. 
\end{abstract}

\section{Light Curve Analysis of CI Aql and Turn-off Time}

     The second recorded outburst of CI Aquilae was discovered 
in 2000 April by Takamizawa since the first recorded outburst in 1917.
After optical brightness reached its maximum, $m_V \sim 9$, 
in early May of 2000, it rapidly decreased to 
$m_V \sim 13$ in about 50 days.  A plateau phase follows;
the brightness leveled off at $m_V \sim 13.5$. 
Once it rapidly decayed to $m_V \sim 14.5$ at the end of November of 2000, 
it stayed at $m_V \sim 14.5$ until March of 2001.  
CI Aql has entered the final decline phase toward its quiescent level
from May of 2001 as shown in Fig. 1
\par
     We have modeled the system consisting of a massive white dwarf (WD)
and a lobe-filling main-sequence (MS) star.  Irradiation effects of 
the accretion disk (ACDK) and the MS companion by the WD are included 
into the light curve calculation.  The numerical method has been 
described in Hachisu \& Kato (2001b).  We are able to reproduce 
the light curve by adopting model parameters similar to those 
of U Sco (Hachisu et al. 2000).  The model parameters including 
those for the ACDK are shown in Fig. 1 (see also Hachisu \& Kato
2001a).
\par
     After the paper of Hachisu \& Kato (2001a) has been published, 
we have new data on the final decay phase of the 2000 
outburst (Schaefer 2001, private communication) and, 
now, are able to determine the turn-off time, 
the helium content of the envelope, and the duration of luminous 
supersoft X-ray phase as shown in Fig. 1.
We have also revised the physical parameters:
the unheated surface temperatures are $T_{\rm ph, MS}= 7100$ K 
of the MS companion, $T_{\rm ph, disk}= 6600$ K of the disk rim,
the apparent distance modulus $(m-M)_V= 13.4$,
the color excess $E(B-V)= 1.0$, the absorption $A_V= 3.1$,
and the distance to CI Aql $d= 1.1$ kpc.

\begin{figure}
\plotone{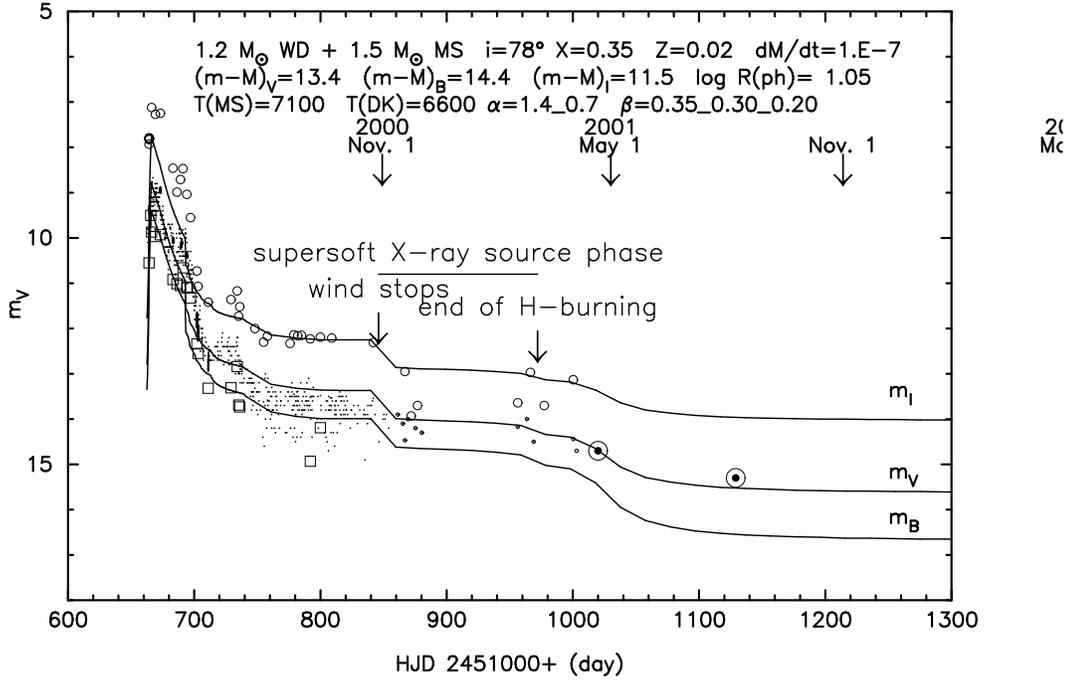}
\caption{
Calculated $V$, $B$, and $I_c$ light curves plotted 
against time (HJD 2,451,000+) together with the observations.
Small dots indicate observational $V$ and visual magnitudes
including late phase $\odot$ marks (Schaefer 2001, private
communication), 
while open squares represent observational $B$ magnitudes
and open circles indicate observational $I_c$ magnitudes 
(all taken from the VSNET archives except $\odot$'s).
Calculated light curves are plotted for the model consisting of 
a $1.2 M_\odot$ WD and $1.5 M_\odot$ MS.  The hydrogen content
of the WD envelope is about $X=0.35$ in mass weight.
Each light curve connects the brightness
at the binary phase 0.35 (roughly the brightest in a binary phase).
The apparent distance modulus of $(m-M)_V= 13.4$ is obtained by fitting.
}
\end{figure}

\end{document}